# THERMODYNAMICS OF MIXTURES CONTAINING AMINES. XVII. $H_m^E$ and $V_m^E$ MEASUREMENTS FOR BENZYLAMINE + HEPTANE OR + 1-ALKANOL MIXTURES AT 298.15 K. APPLICATION OF THE DISQUAC AND ERAS MODELS

Luis Felipe Sanz, Juan Antonio González,[*] Fernando Hevia, Isaías. García de la Fuente, and José Carlos Cobos

[a]G.E.T.E.F., Departamento de Física Aplicada, Facultad de Ciencias, Universidad de Valladolid, Paseo de Belén, 7, 47011 Valladolid, Spain.

*corresponding author, e-mail: jagl@termo.uva.es; Fax: +34-983-423136; Tel: +34-983-423757

**ABSTRACT**

Excess molar enthalpies, $H_m^E$, at 298.15 K and 0.1 MPa have been measured by means of a Tian-Calvet microcalorimeter for the systems benzylamine (phenylmethanamine) + heptane, or + methanol, or + 1-propanol, or + 1-pentanol, or + 1-heptanol, or + 1-decanol. In addition, excess molar volumes, $V_m^E$, at the same conditions have been also determined using a densimeter Anton Paar model DSA 5000 for the benzylamine + heptane mixture. The $H_m^E$ of this solution is large and positive since at 298.15 K the system temperature is close to its upper critical solution temperature. Thus, systems with $n$-alkanes show positive deviations from the Raoult's law. The measured $|V_m^E|$ values are low, indicating the existence of large structural effects. $H_m^E$ values of mixtures involving 1-alkanols are large and negative. That is, interactions between unlike molecules are dominant and the systems are characterized by negative deviations from Raoult's law. It is shown that the enthalpy of the hydrogen bonding between molecules of 1-alkanol and benzyalmine are more negative than those between 1-alkanol molecules. The $V_m^E$ values of the systems with 1-alkanols are also large and negative, and are determined mainly by interactional effects since they increase in line with $H_m^E$ and with the alcohol size. The different contributions to $H_m^E$ have been evaluated. The systems have been studied using the DISQUAC and ERAS models. ERAS describes correctly the $V_m^E$ function. DISQUAC largely improves ERAS results on $H_m^E$ or on excess molar heat capacities at constant pressure for the mixtures with 1-alkanols, which underlines that physical interactions are very relevant in such solutions.

## l. Introduction

Along the last years, we have developed a systematic research on 1-alkanol + amine mixtures. In fact, this type of solutions is very interesting since their behaviour is very different depending on the considered amine. Thus, 1-alkanol + linear primary or secondary amine, or + cyclohexylamine systems show large negative deviations from the Raoult's law [1-4], while the corresponding solutions with aniline [1,5] are characterized by positive deviations from the Raoult's law. In addition, structural effects may be highly relevant as, e.g., in 1-alkanol + *N,N,N*-triethylamine mixtures [4], or in the methanol + aniline solution [1]. Up to now, we have provided volumetric [4,6-12], calorimetric [1,13], phase equilibria [3], viscosimetric [9-11] or dielectric [11,12,14-17] data for this type of solutions, including those with benzylamine [18-20]. Simultaneously, theoretical studies have been conducted [l, 2,4-12,14-17,21-23] by means of several models such as: DISQUAC [24], ERAS [25], UNIFAC [26], Flory [27], the concentration-concentration structure factor [28], the formalism of the Kirkwood-Buff integrals [29,30], or the Kirkwood-Fröhlich's theory [31]. As continuation of this effort, we provide now excess molar enthalpies, $H_m^E$, for methanol or 1-propanol or 1-pentanol, or 1-heptanol or 1-decanol + benzylamine systems at 298.15 K and 0.1 MPa, which are also investigated using the DISQUAC and ERAS models. In order to attain a complete characterization of benzylamine in terms of ERAS, we also report $H_m^E$ and $V_m^E$ (excess molar volume) data for the benzylamine + heptane mixture also at 298.15 K and 0.1 MPa.

Benzylamine and its derived chemicals are important since their use has benefits on human health and safety and for the environment in comparison with benzene which has carcinogenic effects. lt is to be noted that benzylamine is not included as a human carcinogen by the International Agency for Research on Cancer. In addition, benzylamine is biodegradable [32]. All this facilitates its use in technical applications, such as pharmaceuticals, cosmetics, surface active reagents, corrosion inhíbitors, antiseptics or in antimicrobial agents [33-35]. Solutions including benzylamine are also used for $CO_2$ capture [36,37].

## 2. Experimental

### 2.1 Materials

All the chemicals were used as received, without any further purification. Table

1 lists the CAS number of the pure compounds, their source, purity, according to gas chromatographic analysis (GC) provided by the supplier and water contents. In addition, Table 1 also contains a comparison between densities, $\rho$, of the pure compounds and values available in the literature.

*2.2 Apparatus and procedure*

Compounds were weighted using an analytical balance A and D instrument model HR-202 (weighing accuracy 0.1 mg), taking into account the corresponding corrections on buoyancy effects. The standard uncertainty in the final mole fraction is 0.0005. Molar quantities were calculated using the relative atomic mass Table of 2015 issued by the Commission on Isotopic Abundances and Atomic Weights (IUPAC) [38]. All the measurements were carried out at 298.15 K and 0.1 MPa. Calorimetric data were obtained by means of a standard Tian-Calvet microcalorimeter with a temperature stability of 0.01 K. The mixing cell, designed by us, is of aluminium and has a small (< 2%) gas phase. Details on the mixing process and on the calibration of the apparatus have been given previously [1,39]. The estimated maximum standard relative uncertainty for $H_m^E$ is 0.015.

Pt-100 resistances, calibrated using the triple point of water and the melting point of Ga, were used to measure the temperature of the samples, with a standard uncertainty of 0.01 K for $\rho$ measurements. Densities of the pure compounds and of the benzylamine + heptane mixtures were determined using a densimeter and sound analyser Anton Paar, model DSA 5000 with a temperature stability of 0.001 K. More details about the technique, calibration and test of the apparatus can be found elsewhere [12,40]. The standard relative uncertainties of $\rho$ and $V_m^E$ data are estimated to be, respectively, 0.0008 and 0.010.

### 3. Experimental results

Our measurements on $H_m^E$ and $V_m^E$ for the benzylamine + heptane system at 298.15 K and 0.1 MPa are listed in Table 2. Similarly, Table 3 contains $H_m^E$ results for 1-alkanol + benzylamine mixtures at the same conditions. No data have been found in the literature for comparison.

The data ($F_m^E = H_m^E; V_m^E$) were fitted by unweighted least-squares polynomial regression to the equation of the Redlich-Kister type [41]:

$$F_m^E = x_1(1-x_1)\sum_{i=0}^{k-1} A_i (2x_1-1)^i \qquad (1)$$

The number, $k$, of needed coefficients for this regression was determined, for each system, by applying an F-test of additional term [42] at 99.5% confidence level. Tables 4 and 5 list the parameters $A_i$ obtained in the regression, together with the standard deviations $\sigma(F_m^E)$ defined by:

$$\sigma\left(F_m^E\right) = \left[\frac{1}{N-k}\sum_{j=1}^{N}\left(F_{mcal,j}^E - F_{mexp,j}^E\right)^2\right]^{1/2} \quad (2)$$

where $N$ stands for the number of data points, and $F_{mcal,j}^E$ is the value of the excess property calculated using equation (1). Figures S1-S3 (supplementary material) show our experimental values of the excess molar functions together with the results from the corresponding fittings by means of equation (1).

## 4. Models
### 4.1 DISQUAC

DISQUAC is based on the rigid lattice theory developed by Guggenheim [43]. Some important features of the model are summarized. (i) The geometrical parameters: total molecular volumes, $r_i$, surfaces, $q_i$, and the molecular surface fractions, $\alpha_{si}$, of the mixture components are calculated additively using the group volumes $R_G$ and surfaces $Q_G$ recommended by Bondi [44], with the volume $R_{CH4}$ and surface $Q_{CH4}$ of methane taken arbitrarily as volume and surface units [45]. For the groups involved in this investigation, the geometrical parameters are available in the literature [45-47]. (ii) The partition function is factorized into two terms. The excess functions $G_m^E$ and $H_m^E$ are the result of two contributions. The dispersive (DIS) term arises from the contribution of dispersive forces; while the quasichemical (QUAC) term is due to the anisotropy of the field forces created by the solution molecules. The Flory-Huggins equation is used to represent the combinatorial contribution, $G_m^{E,COMB}$, to $G_m^E$ [45,48]. Therefore,

$$G_m^E = G_m^{E,DIS} + G_m^{E,QUAC} + G_m^{E,COMB} \quad (3)$$

$$H_m^E = H_m^{E,DIS} + H_m^{E,QUAC} \quad (4)$$

(iii) The interaction parameters are assumed to depend on the molecular structure of the mixture components; (iv) The same value of the coordination number ($z = 4$) is used for all the polar contacts. This is an important shortcoming of the model, partially removed assuming that the interaction parameters are dependent on the molecular structure. (v) It is also assumed that no change in volume is produced upon mixing (i.e., $V_m^E = 0$).

The equations used to calculate the DIS and QUAC contributions to $G_m^E$ and $H_m^E$ can be found elsewhere [2,49]. The temperature dependence of the interaction parameters is expressed in terms of the DIS and QUAC interchange coefficients [2,49], $C_{st,l}^{DIS}; C_{st,l}^{QUAC}$ where s ≠ t are two contact surfaces present in the mixture and $l$ = 1 (Gibbs energy; $C_{st,1}^{DIS/QUAC} = g_{st}^{DIS/QUAC}(T_o)/RT_o$); $l$ = 2 (enthalpy, $C_{st,2}^{DIS/QUAC} = h_{st}^{DIS/QUAC}(T_o)/RT_o$), $l$ = 3 (heat capacity, $C_{st,3}^{DIS/QUAC} = c_{pst}^{DIS/QUAC}(T_o)/R$)). $T_o$ = 298.15 K is the scaling temperature and $R$, the gas constant.

### 4.2 ERAS

The Extended Real Associated Solution (ERAS) model [25] combines the Real Association Solution Model [50,51] with Flory's equation of state [27]. Some relevant features of the model are the following. (i) The excess molar functions ($F_m^E = H_m^E, V_m^E$) are the result of two contributions. The chemical contribution, $F_{m,chem}^E$, is linked to hydrogen bonding; the physical contribution, $F_{m,phys}^E$, arises from nonpolar Van der Waals interactions and free volume effects. Expressions for $H_m^E$ and $V_m^E$ can be found in previous works [2]. (ii) It is assumed that only consecutive linear association occurs. Thus, self-association is described by a chemical equilibrium constant ($K_i$) independent of the chain length of the self-associated species A or B. In the case of 1-alkanol + benzylamine systems, i = A (1-alkanol) or = B (benzylamine), and the self-association of the compounds is represented by the equations:

$$A_m + A \xleftrightarrow{K_A} A_{m+1} \quad (5)$$

$$B_n + B \xleftrightarrow{K_B} B_{n+1} \quad (6)$$

with m and n ranging from 1 to ∞. The cross-association between two self-associated species $A_m$ and $B_n$ is described by:

$$A_m + B_n \xleftrightarrow{K_{AB}} A_m B_n \quad (7)$$

The cross-association constants, $K_{AB}$, are also assumed that are independent of the chain length. Reactions described by equations (5)-(7) are characterized, respectively, by the molar enthalpies of intermolecular hydrogen-bonding $\Delta h_A^*, \Delta h_B^*$ and $\Delta h_{AB}^*$, and by negative molar hydrogen-bonding volumes, $\Delta v_A^*$, $\Delta v_B^*$ and $\Delta v_{AB}^*$, which are needed to take into account the decrease of the core volume of the molecules upon multimer formation. The three equilibrium constants depend on temperature according to the $\Delta h_i^*$ values and the Van't Hoff equation. (iii) The $F_{m,phys}^E$ term is obtained from the Flory's equation of state [27], which is assumed to be valid for pure compounds and for mixtures [52,53]:

$$\frac{\bar{p}_i \bar{V}_i}{\bar{T}_i} = \frac{\bar{V}_i^{1/3}}{\bar{V}_i^{1/3} - 1} - \frac{1}{\bar{V}_i \bar{T}_i} \qquad (8)$$

where $i$ = A, B or M (mixture). In equation (8), $\bar{V}_i = V_{m,i}/V_{m,i}^*$ ; $\bar{p}_i = p/p_i^*$ ; $\bar{T}_i = T/T_i^*$ are the reduced properties for volume, pressure and temperature, respectively. The pure component reduction parameters ($V_{m,i}^*, p_i^*, T_i^*$) are obtained from $p$-$V$-$T$ data ($\rho$, $\alpha_{pi}$, (isobaric thermal expansión coefficient) and $\kappa_{T_i}$ (isothermal compressibility) and association parameters [52,53]. The reduction parameters for the mixture $p_M^*$ and $T_M^*$ are calculated from mixing rules [52,53]. The total relative molecular volumes and surfaces of the compounds were calculated additively on the basis of the group volumes and surfaces recommended by Bondi [44].

### 5. Adjustment of interaction parameters

#### 5.1 DISQUAC

In terms of DISQUAC, the following surfaces may be present in the investigated systems: (i) type a, aliphatic ($CH_3$, $CH_2$, in 1-alkanols, benzylamine, or $n$-alkanes); (ii) type b, aromatic ($C_6H_5$- in benzylamine or $C_6H_6$ in benzene); (iii) type h, OH in 1-alkanols; (iv) type n, amine ($NH_2$ in benzylamine).

##### 5.1.1 Benzylamine + n-alkane or + benzene

These mixtures are built by three surfaces, a, b and n, which generate three contacts: (a,b), (a,n) and (b,n), The (a,b) contact is represented by dispersive interaction parameters obtained from the study of alkylbenzene (in this case ethylbenzene) + $n$-alkane mixtures [47]. The (a,n) and (b,n) contacts are described by both DIS and QUAC interaction parameters, which must be fitted simultaneously using $H_m^E$ data for the benzylamine + benzene [54] system and liquid-liquid equilibria (LLE) [55] and calorimetric data for the $n$-alkane solutions (this work, [19]). Due to the lack of experimental data, and for the sake of simplicity, we have assumed that the interchange coefficients for $l$ = 1 (Gibbs energy) of the (b,n) contacts are the same that for aniline mixtures [5]. Final values of the $C_{sn,1}^{DIS}$ and $C_{sn,1}^{QUAC}$ (s = a,b; l =1,2,3) coefficients are listed in Table 6.

*5.1.2 1-alkanol + benzylamine*

These systems are built by four surfaces, a,b,h,n, which generate six contacts: (a,b), (a,h), (a,n), (b,h), (b,n) and (h,n). The interaction parameters for the contacts (a,n) and (b,n) have been determined above. The interaction parameters for the contacts (a,h) and (b,h) are known from our DISQUAC studies on 1-alkanol + *n*-alkane [46] or + alkylbenzene [56] systems, respectively. For the (a,b) contacts, the interaction parameters are equal to those used in the preceding subsection. Therefore, only the interaction parameters for the (h,n) contacts must be determined. The general procedure applied in the estimation of the interaction parameters has been explained in detail elsewhere [2,49]. However, the task is now somewhat difficult since no data on vapour-liquid equilibria (VLE) are available for these mixtures. The $C_{nh,1}^{QUAC}$ (l =1,2) were adjusted together with the $C_{hn,2}^{DIS}$ coefficients to get a good description of the symmetry of the $H_m^E$ curves [49]. The first DIS Gibbs energy parameters were then estimated taking into account VLE data for other 1-alkanol + amine systems. Final parameters are listed in Table 6.

*5.2 ERAS*

The values of the ERAS parameters $K_A$, $\Delta h_A^*$, $\Delta v_A^*$ for 1-alkanols, and of their reduction parameters are known from the study of 1-alkanol + *n*-alkane mixtures [25,53,57]. The values of $K_A$, $\Delta h_A^*$, and $\Delta v_A^*$ of benzylamine (Table 7) have been determined using the data reported in this work for the system with heptane. The reduction parameters of benzyalmine (Table 7) were calculated using values of $\alpha_{pi}$, and $\kappa_{T_i}$ from reference [18]. The reduction parameters of heptane are available elsewhere [58]. 1-Alkanol + benzylamine systems are characterized by the binary parameters $K_{AB}$, $\Delta h_{AB}^*$, $\Delta v_{AB}^*$ and $X_{AB}$ (Table 8), which are fitted to the $H_m^E$ and $V_m^E$ data [18]. For the mixtures benzylamine, + heptane, or + benzene, $K_{AB} = \Delta h_{AB}^* = \Delta v_{AB}^* = 0$ and only the $X_{AB}$ values must be fitted (Table 8). It is known that the enthalpy of vaporization of a pure compound, in the framework of the ERAS model, can be determined from the $K_A$, $\Delta h_A^*$, $\Delta v_A^*$ values [59]. Thus, at 298.15 K, ERAS yields 60.7 kJ mol$^{-1}$ for the enthalpy of vaporization of benzylamine, in excellent agreement with the experimental result (60.16 kJ mol$^{-1}$ [60]).

## 6. Theoretical results

DISQUAC results on LLE of benzyalmine + *n*-alkane systems are shown in Table 9 and Figure S4. The model describes correctly the coordinates of the critical points, compositions and upper critical solution temperatures (UCSTs) (Table 9). The coexistence curves are much rounded than the experimental ones (Figure S4) due to DISQUAC is a mean field theory and provides LLE curves which are too high at the UCST and too low at the LCST (lower critical solution temperature) [61-63].

Table 10 contains comparisons between experimental $H_m^E$ data and theoretical results using DISQUAC and ERAS models (see also Figures 1-3). DISQUAC improves results from ERAS for mixtures involving 1-alkanols. Larger differences between ERAS calculations and experimental values emerge for systems with 1-heptanol or 1-decanol (Table 10, Figures 2,3). It is remarkable that ERAS correctly describes the $V_m^E$ curves (see Figure 4). Finally, Table 11 compares DISQUAC results on $C_{pm}^E$ with experimental results for the investigated systems. Figures 5 and 6 show such comparison including ERAS results for the mixtures with heptane or methanol. DISQUAC also improves ERAS calculations for the latter system.

## 7. Discussion

Below, we are referring to values of the thermodynamic properties at equimolar composition and 298.15 K. The number of C atoms of the 1-alkanol is represented by $n_{OH}$.

*7.1 Benzylamine + n-alkane, or +benzene*

Systems with *n*-alkanes are characterized by the antipathy between unlike molecules, i.e., by positive deviations from the Raoult's law. This is supported by the large and positive $H_m^E$ result of the heptane mixture (1778 J mol$^{-1}$, this work) and by the existence of LLE curves with upper critical solution temperatures close to 298.15 K (e.g, 280.1 K for the mixture with decane [55], Table 9), which, in addition, suggests that interactions between benzylamine molecules are mainly of dipolar type. We note that the $H_m^E$ curve is flattened at the top (Figure 1) and that the concentration dependence of $C_{pm}^E$ is W shaped (Figure 5, [19]). Similar features have been also encountered for systems at temperatures close to the UCST [64-66]. The flattening of the $H_m^E$ curves is encountered, e.g., in the mixtures 2,5,8,11-tetraoxadodecane +

dodecane [66] (UCST = 280.81 K, [67]) or dimethyl carbonate + heptane [68] (UCST = 272.7 K, value estimated from experimental data provided in [69]), or butyric anhydride + heptane [70] (UCST = 237.5 K, DISQUAC result), or propanone + heptane (245.2 K, [71]). The experimental data show that the $C_{p\mathrm{m}}^\mathrm{E}$ curves are W-shaped for the mixtures 2,5,8,11-tetraoxadodecane + dodecane [72,73] or dimethyl carbonate + n-alkane [74,75], or acetone + alkane [64]. On the other hand, the $V_\mathrm{m}^\mathrm{E}$ curve of the benzylamine + heptane system, is S-shaped and the absolute value of this excess function at equimolar composition is low (Figure S2), which contrasts with the large and positive $H_\mathrm{m}^\mathrm{E}$ result. This reveals the existence of structural effects. Taking into account the large difference between $\alpha_{p\mathrm{i}}$, values of the mixture components (0.887 $10^{-3}$ $K^{-1}$ for benzylamine [18] and 1.256 $10^{-3}$ $K^{-1}$ for heptane [76]), the mentioned effects can be of free volume type.

It is interesting to compare thermodynamic properties of systems containing aniline, benzylamine or N-methylaniline (NMA). Thus, UCST (aniline + heptane) = 343. I K [77] and UCST (tetradecane)/K = 291.96 (benzylamine) [55], 286.85 (NMA) [78]. That is, interactions between amine molecules become weaker in the sequence: aniline > benzylamine > NMA. The excess molar enthalpies at infinite dilution of the amine, $H_{\mathrm{m1}}^{\mathrm{E},\infty}$, in amine + heptane solutions are (in kJ $\mathrm{mol}^{-1}$): 15 (aniline) [79] and 10 (for both benzylamine (this work, see below) or NMA [78]).

The concentration-concentration structure factor, $S_{\mathrm{CC}}(0)$, [28,80] is a magnitude that allows to investigate the degree of non-randomness in a given system. The ideal mixture is characterized by $S_{\mathrm{CC}}(0)$ = 0.25. Interactions between like molecules (homocoordination) are dominant when $S_{\mathrm{CC}}(0)$ > 0.25. Interactions between unlike molecules are prevalent when $S_{\mathrm{CC}}(0)$ < 0.25. We have evaluated $S_{\mathrm{CC}}(0)$ for the benzylamine or NMA + heptane mixtures and the results are, respectively, 1.436 and 0.777 (Figure S5). Thus, homocoordination is much higher in the former system, since this solution at 298.15 K is closer to its UCST. In fact, W-shaped $C_{p\mathrm{m}}^\mathrm{E}$ curves together with large values of $S_{\mathrm{CC}}(0)$ have been considered as a manifestation of non-randomness effects [73,80,81] which can be detected even above 100 K from the UCST [81].

For systems containing benzene, $H_\mathrm{m}^\mathrm{E}$/J $\mathrm{mol}^{-1}$ changes in the order: 751 (aniline) [82] > 490 (benzylamine, T= 303.15 K) [54] > 457 (NMA) [83], which may be ascribed to aniline-aniline interactions are stronger.

### 7.2. 1-alkanol + benzylamine

These systems show large and negative $H_m^E$ values, particulary when the shorter 1-alkanols are involved (Table 10, Figure 7a) and, therefore, are characterized by interactions between unlike molecules (negative deviations from Raoult's law). The $V_m^E$ values are also large and negative: $-1.542$ ($n_{OH} = 1$); $-1.034$ ($n_{OH} = 3$); $-0.765$ ($n_{OH} = 5$) [18]. We note that $V_m^E$ and $H_m^E$ results change in line and increase when $n_{OH}$ is increased. Consequently, the main contribution to $V_m^E$ arises from interactional effects. Systems formed by 1-alkanol and 1-hexylamine [4,52], or cyclohexylamine [l], or di-$n$-propylamine [6] show similar trends. The $C_{pm}^E$ values of the studied mixtures are large and positive, a typical feature of solutions where association/solvation effects are determinant when describing their thermodynamic properties. Thus, $C_{pm}^E$ (heptane)/J mol$^{-1}$ K$^{-1}$ = 11.7 (ethanol); 11.1 (1-butanol) [84] and $C_{pm}^E$/J mol$^{-1}$ K$^{-1}$ = 13.4 (1-octanol + di-$n$-ethylamine) [85] or $C_{pm}^E$ (benzylamine)/J mol$^{-1}$ K$^{-1}$ = 12.9 (methanol) [19]; 12 (1-butanol) [20]. These values suggest that solvation effects contribute more largely to $C_{pm}^E$ than those related to alcohol self-association. On the other hand, both the $H_m^E$ and $C_{pm}^E$ curves are slightly shifted to higher 1-alkanol concentrations (Figures 2,3 and 6) which may be ascribed to solvation effects become more relevant at the mentioned concentrations.

Now, the enthalpy of the 1-alkanol-benzylamine interactions (termed $\Delta H_{OH-NH2}^{int}$) is evaluated. Neglecting structural effects [61,86], it is possible to assume that $H_m^E$ is the result of three contributions. The positive contributions $\Delta H_{OH-OH}, \Delta H_{NH2-NH2}$ are related, respectively, to the breaking of the alkanol and amine networks along the mixing process. When the mixture takes place, new OH---NH2 interactions are created, and it implies a negative contribution, $\Delta H_{OH-NH2}$, to $H_m^E$. Thus [87-89]:

$$H_m^E = \Delta H_{OH-OH} + \Delta H_{NH2-NH2} + \Delta H_{OH-NH2} \qquad (9)$$

The values of $\Delta H_{OH-NH2}^{int}$ can be obtained extending the equation (9) to $x_1 \to 0$ [89-91]. Then, $\Delta H_{OH-OH}$ and $\Delta H_{NH2-NH2}$ can be replaced by $H_{m1}^{E,\infty}$ (partial excess molar enthalpy at infinite dilution of the first component) of 1-alkanol or benzylamine + $n$-alkane systems. In such a case,

$$\Delta H_{OH-NH2}^{int} = H_{m1}^{E,\infty}(1-\text{alkanol} + \text{benzylamine})$$

$$-H_{m1}^{E,\infty}(1-\text{alkanol} + \text{heptane}) - H_{m1}^{E,\infty}(\text{benzylamine} + \text{heptane}) \quad (10)$$

For 1-alkanol + *n*-alkane systems, we have taken $H_{m1}^{E,\infty}$ = 23.2 kJ·mol$^{-1}$ [92-94], the same value for all the 1-alkanols, which is a common approach within association theories [25,95-97]. For the systems 1-alkanol + benzylamine or benzylamine + heptane, $H_{m1}^{E,\infty}$ data were determined from $H_m^E$ measurements over the entire mole fraction range. A similar procedure was applied when we determined the enthalpy of interactions between 1-alkanol and different organic solvents, such linear monoether [89] or polyether [98], or linear alkanone [99], or nitrile [100], or nitroalkane [101] or linear organic carbonate [l02]. This makes that $\Delta H_{\text{OH-NH2}}^{\text{int}}$ values collected in Table 12 (Figure 7b) are meaningful. Inspection of the mentioned Table shows that interactions between unlike molecules are stronger in the methanol system, and weaker in the mixture with 1-decanol. For the remainder solutions, $\Delta H_{\text{OH-NH2}}^{\text{int}}$ decreases smoothly. lt is remarkable that $\Delta H_{\text{OH-NH2}}^{\text{int}}$ values are lower than those corresponding to the alkanol-alkanol interactions ($-$23.2 kJ mol$^{-1}$). That is, alkanol-amine interactions are stronger than those between alkanol molecules. This result has been also encountered for other 1-alkanol + amine systems [52,53,57]. We investigate now the variation with $n_{\text{OH}}$ of the different contributions to $H_m^E$

### 7.2.1 The $\Delta H_{\text{OH-OH}}$ term

For mixtures involving ethylbenzene (isomeric molecule of benzylamine), $H_m^E$/J mol$^{-1}$ = 684 ($n_{\text{OH}}$ = 1); 908; ($n_{\text{OH}}$ = 3); 932 ($n_{\text{OH}}$ = 5); 895 ($n_{\text{OH}}$ = 8) [103]. Such values suggest that the positive $\Delta H_{\text{OH-OH}}$ contribution remains nearly constant for larger 1-alkanols.

### 7.2.2 The $\Delta H_{\text{NH2-NH2}}$ term

This is positive contribution to $H_m^E$ increases in line with $n_{\text{OH}}$ since the larger aliphatic surfaces of longer 1-alkanols break more easily the interactions between benzyalmine molecules. Note that the UCSTs of the benzylamine + *n*-alkane systems increase with the number of C atoms in the alkane (Table 9).

### 7.2.3 The $\Delta H_{\text{OH-NH2}}$ term

This term contributes negatively to $H_m^E$ and increases with $n_{\text{OH}}$ due to: (i) interactions between unlike molecules become weaker at the mentioned condition (Table 12); and (ii) the creation of interactions between unlike molecules is less probable since the OH group is more sterically hindered in longer 1-alkanols.

The increase of the $H_m^E$ (Figure 7a) values with $n_{\text{OH}}$ may be now explained taking into account that the contributions $\Delta H_{\text{NH2-NH2}}$ and $\Delta H_{\text{OH-NH2}}$ increase in line with $n_{\text{OH}}$ while the term $\Delta H_{\text{OH-OH}}$ is more or less constant for larger $n_{\text{OH}}$ values. In the temperature range (298.15-308.15) K, $C_{pm}^{\text{E}}$ slightly increases for the system with methanol, and slightly decreases for the remainder solutions (Table 11), which is probably due to solvation effects are more relevant in the former mixture. At higher temperatures, e.g., 333.15 K, the $C_{pm}^{\text{E}}$ / J mol$^{-1}$ K$^{-1}$ values become lower and change in the order: 9.9 (1-butanol) > 7.0 (1-octanol) > 6.2 (1-decanol) [20]. Such variation may be explained in terms of large alcohol dissociation produced at high temperatures [104], and to a weakening of alkanol-amine interactions (lower solvation effects) (Table 12). A similar temperature dependence of $C_{pm}^{\text{E}}$ has been also observed in the systems 1-butanol + toluene [104], or 1-alkanol + acetophenone [105,106], or + dimethylsulfoxide [107].

Finally, we show in Figure 7b a comparison between $\Delta H_{\text{OH-NH2}}^{\text{int}}$ values for 1-alkanol + benzylamine, or + aniline, or + NMA mixtures. Similarly, the corresponding $H_{\text{m}}^{\text{E}}$ results are compared in Figure 7a. We note that, for a given 1-alkanol, interactions between unlike molecules become stronger in the sequence: NMA < aniline < benzylamine. This, together with the fact that the amine group is more sterically hindered in NMA, explains that $H_{\text{m}}^{\text{E}}$ changes in the order: NMA > aniline > benzylamine and that the $H_{\text{m}}^{\text{E}}$ variation is sharper in mixtures containing NMA.

*7.3 Excess molar internal energies at constant volume*

This magnitude, represented by $U_{\text{m},V}^{\text{E}}$, can be determined from [61]:

$$U_{\text{m},V}^{\text{E}} = H_{\text{m}}^{\text{E}} - T\frac{\alpha_p}{\kappa_T}V_{\text{m}}^{\text{E}} \qquad (11)$$

where $\alpha_p$ and $\kappa_T$ are, respectively, the isobaric thermal expansion coefficient and the coefficient of isothermal compressibility of the considered system. The $T\frac{\alpha_p}{\kappa_T}V_{\text{m}}^{\text{E}}$ is the equation of state (eos) contribution to $H_{\text{m}}^{\text{E}}$. Here, the $\alpha_p$ and $\kappa_T$ values were calculated assuming ideal behavior for the systems ($M^{\text{id}} = \phi_1 M_1 + \phi_2 M_2$; with $M_i = \alpha_{pi}$, or $\kappa_{T_i}$ and $\phi_i = x_i V_{\text{m},i}/(x_1 V_{\text{m},1} + x_2 V_{\text{m},2})$). For pure compounds, their $\alpha_{pi}$, and $\kappa_{T_i}$ values were taken from the literature [89,108]. For heptane mixtures, $U_{\text{m},V}^{\text{E}}$ /J mol$^{-1}$ = 1743 (NMA) and 1784 (benzylamine). These similar results reveal that more interactions between NMA molecules are broken upon mixing which is in agreement with the fact that NMA-NMA interactions are weaker. In the case of systems with 1-alkanols, the contributions to $H_m^E$ from the eos term are large an negative. They are 22.7%, 26.4% and 25% for the systems with methanol, 1-propanol or 1-pentanol, respectively. The values of ($U_{m,V}^E$ /J mol$^{-1}$) in the same order as above, are: $-2071$, $-1098$ and $-861$. The $U_{m,V}^E$ variation is smoother than the $H_m^E$ variation, which may be ascribed, at least in part, to the large contribution to $H_m^E$ from the eos term. No $V_m^E$ data have been encountered in the literature to calculate $U_{m,V}^E$ of 1-alkanol + NMA mixtures. However, in a previous work [12], we determined this property for solutions containing aniline and the results were (in J mol$^{-1}$): 143 (methanol); 997 (1-propanol) and 1104 (1-pentanol). The corresponding ($H_m^E$/J mol$^{-1}$) values, in the same sequence, are: $-175$, 776, 1011 [109]. We note that $U_{m,V}^E$ changes more smoothly with the alkanol size and particularly the large contribution from the eos term to $H_m^E$ in the case of the mixture with methanol.

*7.4 The interaction parameters*

*7.4.1 ERAS parameters*

Firstly, we underline that our values of the $K_A$, $\Delta h_A^*$, $\Delta v_A^*$ parameters for benzyalmine are reliable since they conduct to a value of the enthalpy of vaporization in good agreement with the experimental value (see above). Othe other hand, these parameters for aniline [5], benzylamine (this work) and *N*-methylaniline [108] are rather consistent between them. Thus, self-association is higher in aniline or benzylamine ($K_A$ = 14.8) than in NMA ($K_A$ = 6). Amine-amine interactions are stronger between aniline molecules ($\Delta h_A^* = -15$ kJ mol$^{-1}$) since $\Delta h_A^* = -12.5$ kJ mol$^{-1}$ for benzylamine and NMA. The latter result is supported by the similar values of the UCST of systems with tetradecane and benzylamine (290 K [78]) or NMA (286.8 K [55]). In any case, mixtures with alkanes are characterized by large $X_{AB}$ values ([5,78] and Table 8), and this means that physical interactions are determinant in such solutions. Thus, for the benzylamine + heptane mixture we have: $H_{m,phys}^E$ = 1317 J mol$^{-1}$ and $H_{m,chem}^E$ = 485 J mol$^{-1}$. For 1-alkanol + benzylamine mixtures, we note that the $\Delta h_{AB}^*$ values are close to those of $\Delta H_{OH-NH2}^{int}$ (Tables 8 and 12, Figure 7b), which supports our calculations. The physical parameters are low, and thermodynamic properties are mainly determined by association/solvation effects. For example, in the case of the methanol mixture, $H_{m,phys}^E$ = 11 J mol$^{-1}$ and $H_{m,chem}^E$ = $-2690$ J mol$^{-1}$, and for the 1-pentanol system: $H_{m,phys}^E$ = 127 J mol$^{-1}$ and $H_{m,chem}^E$ = $-1218$ J mol$^{-1}$. The poor $H_m^E$ results obtained for 1-alkanols ($\neq$ methanol) remark the importance of dipolar interactions, particularly for systems with the longer 1-alkanols. We also show some results on $C_{pm}^E$ (Figures 5 and 6).

Calculations require an additional parameter [110], $\frac{dX_{AB}}{dT}$ /J cm$^{-3}$ K$^{-1}$ = 0.13 (benzylamine + heptane, $T$ = 293.15 K); 1.0 (methanol + benzylamine). Of course, the W-shaped $C_{pm}^E$ curve of the heptane system is not described by the model (DISQUAC also does not represent this complex dependence with the concentration of $C_{pm}^E$ (Figure 5)) . For the mixture with methanol, $C_{pm}^E$ is shifted towards low mole fractions of the alcohol, which seems to indicate that self-association effects are overestimated.

*7.4.2 DISQUAC interaction parameters.*

The $C_{an,1}^{DIS}$ coefficients slightly change with the *n*-alkane size (Table 6). As already mentioned, DISQUAC provides LLE curves which are too high at the UCST and too low at the LCST, which explains the mentioned variation of the $C_{an,1}^{DIS}$ coefficients in order to provide not very high calculated UCSTs [62,63]. In mixtures including 1-alkanols, the QUAC parameters are independent of the alcohol, a behavior often observed when mixtures with alcohols are investigated in terms of DISQUAC. Here, we remark, that these parameters largely differ from those of systems with aniline [5] or NMA [78], which may be related to the thermodynamic properties of 1-alkanol + aniline or + NMA systems are quite similar between them and that differ from those of the benzyalmine mixtures. Figure 7a illustrates this fact since we can see that the $H_m^E$ values of systems with aniline or NMA are very different to those of benzylamine. Finally, we remark that DISQUAC describes correctly $C_{pm}^E$ and the temperature dependence of this excess function in the case of systems with 1-alkanols (Table 11, Figure 6)). In view of the results, one can conclude that DISQUAC is a useful tool to describe consistently a set of thermodynamic properties of systems, independently of their deviations from the Raoult's law (positive or negative). At this regards, it is remarkable that, e.g., model calculations show that heterocoordination (interactions between unlike molecules) is the dominant trend in the mixture with methanol (Figure S5).

## 8. Conclusions

Measurements on $H_m^E$ at 298.15 K and 0.1 MPa have been are reported for the benzylamine + heptane, or + methanol, or + 1-propanol, or + 1-pentanol, or + 1-heptanol, or + 1-decanol systems. $V_m^E$ data at the same conditions are also given for the benzylamine + heptane mixture. Mixtures with *n*-alkanes are characterized by positive deviations from the Raoult's law. Structural effects exist in the solution with heptane. Systems with 1-alkanols show negative $H_m^E$ values which increase with the alcohol size, and therefore, are characterized by interactions between unlike molecules (negative deviations from the Raoult's law). The $V_m^E$ values of these systems are essentially determined by interactional effects. 1-Alkanol-benzylamine interactions become weaker when the chaing length of the 1-alcohol increases, and are stronger than those between 1-alkanol molecules. It has been shown that the $\Delta H_{\text{NH2-NH2}}$ and $\Delta H_{\text{OH-NH2}}$ contributions increase in line with $n_{\text{OH}}$ while the term $\Delta H_{\text{OH-OH}}$ is nearly constant. The mixtures have been studied using DISQUAC and ERAS. The latter model describes correctly the $V_m^E$ curves. DISQUAC largely improves ERAS results on $H_m^E$ or on $C_{pm}^E$ for systems including 1-alkanols, which reamrks the importance of physical interactions in the investigated solutions.

**Funding**

This work was supported by Consejería de Educación de Castilla y León, under Project VA100G19 (Apoyo a GIR, BDNS: 425389).

(accessed 2020).

**CRediT authorship contribution statement**

L.F. Sanz, Conceptualization, Data Curation, Software, Validation, Original Draft. J.A. González, Conceptualization, Formal Analysis, Methology, Review and Editing. F. Hevia, Data Curation, Formal Analysis, Investigation, Writing. I. García de la Fuente, Investigation, Supervision, Original Draft, Writing. J.C. Cobos, Methodology, Investigation, Supervision, Validation.

**TABLE 1**

Properties of pure compounds: CAS number, source, purity, water content and density, $\rho$, at 298.15 K and 0.1 MPa.

| Compound | CAS | Source | Purity[a] | Water content[a] | $\rho$ [b]/g cm$^{-3}$ | |
|---|---|---|---|---|---|---|
| | | | | | Experimental | Literature |
| heptane | 142-82-5 | Sigma-Aldrich | 0.998 | 4 10$^{-5}$ | 0.679606 | 0.6796 [111,112] 0.6794 [113,114] 0.67978 [115] 0.6794 [116] |
| benzylamine | 100-46-9 | Fluka | 0.998 | 6.8 10$^{-4}$ | 0.97809 | 0.981 [117] 0.978337 [19] 0.97935 [118] |
| methanol | 67-56-1 | Sigma-Aldrich | 0.999 | 2 10$^{-5}$ | 0.78720 | 0.7869 [119] 0.786884 [120] |
| 1-propanol | 71-23-8 | Fluka | 0.999 | 1 10$^{-3}$ | 0.79951 | 0.79960 [76,121] 0.79959 [122] |
| 1-pentanol | 71-41-0 | Sigma-Aldrich | 0.999 | 3 10$^{-4}$ | 0.81087 | 0.81080 [121] 0.81103 [122] |
| 1-heptanol | 111-70-6 | Sigma-Aldrich | 0.998 | 3 10$^{-3}$ | 0.818987 | 0.81875 [123] |
| 1-decanol | 112-30-1 | Sigma-Aldrich | 0.987 | 3 10$^{-3}$ | 0.826581 | 0.82644 [123] |

[a]in mole fraction, by gas chromatography. Initial purity provided by the supplier. [b]in mass fraction (Karl-Fischer titration); [c]standard uncertainties (*u*): $u(T) = 0.01$ K; $u(p) = 10$ kPa; relative standard uncertainty, $u_r(\rho) = 0.0008$

**TABLE 2**

Excess molar volumes, $V_m^E$, and excess molar enthalpies, $H_m^E$, at 298.15 K and 0.1 MPa for the benzylamine(1) + heptane(2) mixture.[a]

| $x_1$ | $V_m^E$ /cm$^3$ mol$^{-1}$ | $x_1$ | $H_m^E$/J mol$^{-1}$ |
|---|---|---|---|
| 0.0500 | 0.0667 | 0.0547 | 505 |
| 0.1003 | 0.1179 | 0.1003 | 823 |
| 0.1468 | 0.1240 | 0.1475 | 1091 |
| 0.1973 | 0.1196 | 0.1991 | 1325 |
| 0.2398 | 0.1008 | 0.3003 | 1625 |
| 0.2996 | 0.0628 | 0.4077 | 1754 |
| 0.3989 | 0.0123 | 0.4972 | 1809 |
| 0.4991 | −0.0218 | 0.5969 | 1678 |
| 0.6010 | −0.0586 | 0.6926 | 1532 |
| 0.7000 | −0.0804 | 0.7935 | 1230 |
| 0.7984 | −0.1027 | 0.8551 | 961 |
| 0.8491 | −0.1113 | 0.9049 | 687 |
| 0.9005 | −0.1127 | 0.9541 | 359 |
| 0.9486 | −0.0789 | | |

[a] Standard uncertainties (*u*): $u(T) = 0.01$ K; $u(p) = 10$ kPa; $u(x_1) = 0.0005$; $u(V_m^E) = 0.010 \cdot |V_m^E|_{max} + 0.005$ cm$^3$·mol$^{-1}$. For $H_m^E$, the relative combined expanded uncertainty (0.95 level of confidence) is $U_{rc}(H_m^E) = 0.03$.

**TABLE 3**

Excess molar enthalpies, $H_m^E$, at 298.15 K and 0.1 MPa for 1-alkanol(1) + benzylamine(2) mixtures.[a]

| $x_1$ | $H_m^E$/J mol$^{-1}$ | $x_1$ | $H_m^E$/J mol$^{-1}$ |
|---|---|---|---|
| methanol(1) + benzylamine(2) | | | |
| 0.0665 | −524 | 0.6009 | −2668 |
| 0.1204 | −926 | 0.7039 | −2293 |
| 0.1785 | −1335 | 0.7665 | −2079 |
| 0.2020 | −1488 | 0.7943 | −1878 |
| 0.2661 | −1874 | 0.8460 | −1486 |
| 0.3112 | −2103 | 0.8929 | −1080 |
| 0.4067 | −2477 | 0.9494 | −547 |
| 0.4921 | −2692 | | |
| 1-propanol(1) + benzylamine(2) | | | |
| 0.0601 | −289 | 0.5971 | −1464 |
| 0.1036 | −484 | 0.6945 | −1265 |
| 0.1555 | −700 | 0.7447 | −1117 |
| 0.1987 | −866 | 0.7998 | −926 |
| 0.2464 | −1029 | 0.8508 | −720 |
| 0.3003 | −1186 | 0.8993 | −499 |
| 0.4068 | −1405 | 0.9488 | −259 |
| 0.5076 | −1491 | | |
| 1-pentanol(1) + benzylamine(2) | | | |
| 0.0542 | −188 | 0.6010 | −1119 |
| 0.0979 | −337 | 0.7053 | −962 |
| 0.1462 | −493 | 0.7514 | −866 |
| 0.2001 | −617 | 0.8058 | −695 |
| 0.2452 | −766 | 0.8487 | −562 |
| 0.2928 | −874 | 0.9009 | −366 |
| 0.4023 | −1072 | 0.9442 | −215 |
| 0.4997 | −1159 | 0.9466 | −202 |
| 1-heptanol(1) + benzylamine(2) | | | |
| 0.0535 | −127 | 0.6007 | −963 |
| 0.1010 | −232 | 0.7036 | −855 |
| 0.1486 | −346 | 0.7488 | −755 |
| 0.1971 | −466 | 0.7974 | −648 |

Table 3 (continued)

| $x_1$ | $H_m^E$ | $x_1$ | $H_m^E$ |
|---|---|---|---|
| 0.2629 | −623 | 0.8531 | −470 |
| 0.2967 | −691 | 0.8972 | −344 |
| 0.3984 | −869 | 0.9392 | −211 |
| 0.5044 | −988 | | |
| | 1-decanol(1) + benzylamine(2) | | |
| 0.0568 | −19 | 0.6032 | −729 |
| 0.1020 | −61 | 0.6994 | −658 |
| 0.1329 | −100 | 0.7400 | −593 |
| 0.1986 | −211 | 0.7867 | −509 |
| 0.2486 | −295 | 0.8444 | −403 |
| 0.3038 | −415 | 0.8857 | −306 |
| 0.3964 | −574 | 0.9388 | −165 |
| 0.4971 | −687 | | |

$^a$The standard uncertainties are: $u(T) = 0.01$ K, $u(p) = 10$ kPa, and $u(x_1) = 0.0005$. The relative combined expanded uncertainty of $H_m^E$ (0.95 level of confidence) is $U_{rc}(H_m^E) = 0.03$

**TABLE 4**

Coefficients $A_i$ and standard deviations, $\sigma(F_m^E)$ (equation (2)), for the representation of $F_m^E (= H_m^E; V_m^E)$ data at 298.15 K and 0.1 MPa of the benzylamine(1) + heptane(2) mixture by equation (1).

| $F_m^E$ | $A_0$ | $A_1$ | $A_2$ | $A_3$ | $A_4$ | $\sigma(F_m^E)$ [a] |
|---|---|---|---|---|---|---|
| $H_m^E$ / J mol$^{-1}$ | 7113 | −689 | 2234 | | | 14 |
| $V_m^E$ /cm$^3$ mol$^{-1}$ | −0.1143 | −0.6480 | −1.3988 | −0.7406 | −0.7406 | 0.005 |

[a] in the same units that $F_m^E$

**TABLE 5**

Coefficients $A_i$ and standard deviations, $\sigma(H_m^E)$ (equation (2)), for the representation of $H_m^E$ data at 298.15 K and 0.1 MPa of 1-alkanol(1) + benzylamine(2) mixtures by equation (1).

| 1-alkanol | $A_0$ | $A_1$ | $A_2$ | $A_3$ | $\sigma(H_m^E)$ / J mol$^{-1}$ |
|---|---|---|---|---|---|
| methanol | −10713 | −1977 | 1106 | 375 | 25 |
| 1-propanol | −5968 | −487 | 977 | 498 | 5.9 |
| 1-pentanol | −4605 | −594 | 986 | 487 | 11 |
| 1-heptanol | −3898 | −1007 | 1202 | 470 | 7.4 |
| 1-decanol | −2769 | −1466 | 1549 | | 6.5 |

**TABLE 6**

Dispersive (DIS) and quasichemical (QUAC) interchange coefficients, $C_{sn,l}^{DIS}$ and $C_{sn,l}^{QUAC}$, for (s,n) contacts (s = a, aliphatic; b, aromatic; h, OH) in benzylamine + organic solvent mixtures mixtures ($l$ = 1, Gibbs energy; $l$ = 2, enthalpy; $l$ = 3, heat capacity)

| Solvent | contact | $C_{sn,1}^{DIS}$ | $C_{sn,2}^{DIS}$ | $C_{sn,3}^{DIS}$ | $C_{sn,1}^{QUAC}$ | $C_{sn,2}^{QUAC}$ | $C_{sn,3}^{QUAC}$ |
|---|---|---|---|---|---|---|---|
| benzene | (b,n) | 3.7 | −5.05 | −4 | 1.25 | 10 | 4 |
| *n*-alkane | (a,n) | 1.65 | −1.25 | 3.2 | 5 | 10 | 1.5 |
| methanol | (h,n) | 1.4[b] | −11 | 21 | −3 | −2 | 7.5 |
| 1-propanol | (h,n) | 1.8[b] | −2.9 | −10 | −3 | −2 | 7.5 |
| 1-butanol | (h,n) | 1.8[b] | −1.7 | −15 | −3 | −2 | 7.5 |
| 1-pentanol | (h,n) | 1.8[b] | −0.5 | −30 | −3 | −2 | 7.5 |
| 1-heptanol | (h,n) | 1.8[b] | 1.5 | −70 | −3 | −2 | 7.5 |
| 1-decanol | (h,n) | 1.8[b] | 5.5 | −90 | −3 | −2 | 7.5 |

[a] $C_{sn,1}^{DIS}$ = 1.6 (dodecane); 1.55 (≥ tetradecane); [b] guessed value

## TABLE 7

Physical properties and ERAS parameters for benzylamine at 298.15 K and 0.1 MPa: $V_m$, molar volume; $\alpha_p$ isobaric thermal expansion coefficient; $\kappa_T$, isothermal compressibility; $K_A$, equilibrium constant of self-association; $\Delta h_A^*$, molar enthalpy of intermolecular hydrogen-bonding between benzylamine molecules; $\Delta v_A^*$, molar hydrogen-bonding volume between benzylamine molecules, $V_m^*$, reduction parameter for volume; $p^*$, reduction parameter for pressure

| Property/parameter | |
|---|---|
| $V_m$/cm$^3$ mol$^{-1}$ | 109.55 |
| $\alpha_p$ /10$^{-3}$ K$^{-1}$ | 0.887 [18] |
| $\kappa_T$ /10$^{-12}$ Pa | 544.5 [18] |
| $K_A$ | 14.8 |
| $\Delta h_A^*$/kJ mol$^{-1}$ | −12.5 |
| $\Delta v_A^*$/ cm$^3$ mol$^{-1}$ | −5 |
| $V_m^*$ /cm$^3$ mol$^{-1}$ | 90.88 |
| $p^*$/ J cm$^{-3}$ | 643.9 |

## TABLE 8

ERAS parameters[a] for benzylamine(A) + benzene(B) or + heptane(B) and for 1-alkanol(A) + benzylamine(B) mixtures at 298.15 K and 0.1 MPa

| System | $K_{AB}$ | $\Delta h_{AB}^*$ / kJ mol$^{-1}$ | $\Delta v_{AB}^*$ / cm$^3$ mol$^{-1}$ | $X_{AB}$ / J mol$^{-3}$ |
|---|---|---|---|---|
| benzylamine + benzene[b] | 0 | 0 | 0 | 11.50 |
| benzylamine + heptane | 0 | 0 | 0 | 50 |
| methanol + benzylamine | 3500 | −46.2 | −13.2 | 2 |
| 1-propanol + benzylamine | 1500 | −38.2 | −11.2 | 4 |
| 1-pentanol + benzylamine | 1000 | −36.7 | −10.9 | 6 |
| 1-heptanol + benzylamine | 600 | −35.5 | −10.7 | 8 |
| 1-decanol + benzylamine | 120 | −35.2 | −10.4 | 9 |

[a] $K_{AB}$, equilibrium constant; $\Delta h_{AB}^*$, molar enthalpies of intermolecular hydrogen-bonding; $\Delta v_{AB}^*$, molar hydrogen-bonding volumes; $X_{AB}$, physical parameter; [b]system at 303.15 K

**TABLE 9**

Critical points, temperatures, $T_c$, and compositions, $x_{1c}$, for benzylamine(1) + n-alkane(2) mixtures. Comparison of experimental results (Exp.) [55] with DISQUAC (DQ) values obtained using interaction parameters from Table 6.

| n-alkane | $T_c$ /K | | $x_{1c}$ | |
|---|---|---|---|---|
| | Exp. | DQ. | Exp. | DQ. |
| n-C$_{10}$ | 280.1 | 283.7 | 0.569 | 0.572 |
| n-C$_{12}$ | 286.8 | 290.5 | 0.632 | 0.630 |
| n-C$_{14}$ | 291.9 | 295 | 0.707 | 0.680 |
| n-C$_{16}$ | 298.3 | 301.3 | 0.745 | 0.722 |

**TABLE 10**

Molar excess enthalpies, $H_m^E$, at equimolar composition, 298.15 K and 0.1 MPa of binary mixtures involving benzylamine. Comparison with ERAS[a] and DISQUAC results obtained using parameters from Tables 6-8.

| Solvent | N | $H_m^E$/ J mol$^{-1}$ | | | $dev(H_m^E)$[b] | | |
|---|---|---|---|---|---|---|---|
| | | Exp. | ERAS | DQ | Exp. | ERAS | DQ |
| Benzene[c] | 13 | 498 | 507 | 510 | 0.013 | 0.141 | 0.022 |
| heptane | 13 | 1778 | 1803 | 1818 | 0.008 | 0.021 | 0.026 |
| methanol | 15 | –2678 | –2680 | –2670 | 0.009 | 0.163 | 0.019 |
| 1-propanol | 15 | –1492 | –1390 | –1525 | 0.039 | 0.152 | 0.033 |
| 1-pentanol | 16 | –1151 | –1091 | –1180 | 0.011 | 0.215 | 0.038 |
| 1-heptanol | 15 | –974 | –807 | –934 | 0.008 | 0.222 | 0.054 |
| 1-decanol | 15 | –692 | –474 | –690 | 0.009 | 0.267 | 0.068 |

[a]In ERAS calculations, benzylamine is the first component in systems with heptane or benzene, and the second component in mixtures including 1-alkanols;

[b] $dev(H_m^E) = \left[ \frac{1}{N} \sum \left( \frac{H_{m,calc}^E - H_{m,exp}^E}{H_{m,exp}^E(x_1 = 0.5)} \right)^2 \right]^{1/2}$ (N, number of data points); [c]system at 303.15 K

**TABLE 11**

Molar heat capacities at constant pressure, $C_{pm}^{E}$ at equimolar composition, 298.15 K and 0.1 MPa of benzylamine + organic solvent mixtures. Comparison of experimental values (Exp.) with DISQUAC (DQ) results obtained using parameters from Table 6.

| System | $T$/K | $C_{pm}^{E}$ / J mol$^{-1}$ K$^{-1}$ | | Ref. |
|---|---|---|---|---|
| | | Exp. | DQ | |
| Benzylamine + heptane | 293.15 | 6.3 | 6.3 | 19 |
| Methanol + benzylamine | 298.15 | 12.9 | 13 | 19 |
| | 308.15 | 13.3 | 14.6 | 19 |
| 1-propanol + benzylamine | 298.15 | 12.7 | 13 | 19 |
| | 308.15 | 12.4 | 13.2 | 19 |
| 1-butanol + benzylamine | 298.15 | 12 | 12 | 20 |
| | 308.15 | 11.6 | 11.8 | 20 |
| | 333.15 | 9.9 | 8.4 | 20 |
| 1-pentanol + benzylamine | 298.15 | 13.0 | 13.1 | 19 |
| | 308.15 | 11.9 | 12.7 | 19 |
| 1-decanol + benzylamine | 298.15 | 11.4 | 11.7 | 20 |
| | 308.15 | 9.9 | 11.1 | 20 |
| | 333.15 | 6.2 | 7.2 | 20 |

**TABLE 12**

Partial molar excess enthalpies at infinite dilution of the first component,[a] $H_{m1}^{E,\infty}$, at 298.15 K at 0.1 MPa for amine(1) + alkane(2) and for 1-alkanol(1) + amine(2) mixtures and hydrogen bond enthalpies, $\Delta H_{OH-NH2}^{int}$, for 1-alkanol-amine interactions.

| System | $H_{m1}^{E,\infty}$ /kJ mol$^{-1}$ | $\Delta H_{OH-NH2}^{int}$ /kJ mol$^{-1}$ |
|---|---|---|
| benzylamine(1) + heptane(2) | 10 | |
| methanol(1) + benzylamine(2) | −7.8 | −41 |
| 1-propanol(1) + benzylamine(2) | −4.5 | −37.7 |
| 1-pentanol(1) + benzylamine (2) | −3.5 | −36.7 |
| 1-heptanol(1) + benzylamine (2) | −2.2 | −35.4 |
| 1-decanol(1) + benzylamine e(2) | 0.2 | −33 |

[a]values obtained from $H_m^E$ data over the whole concentration range

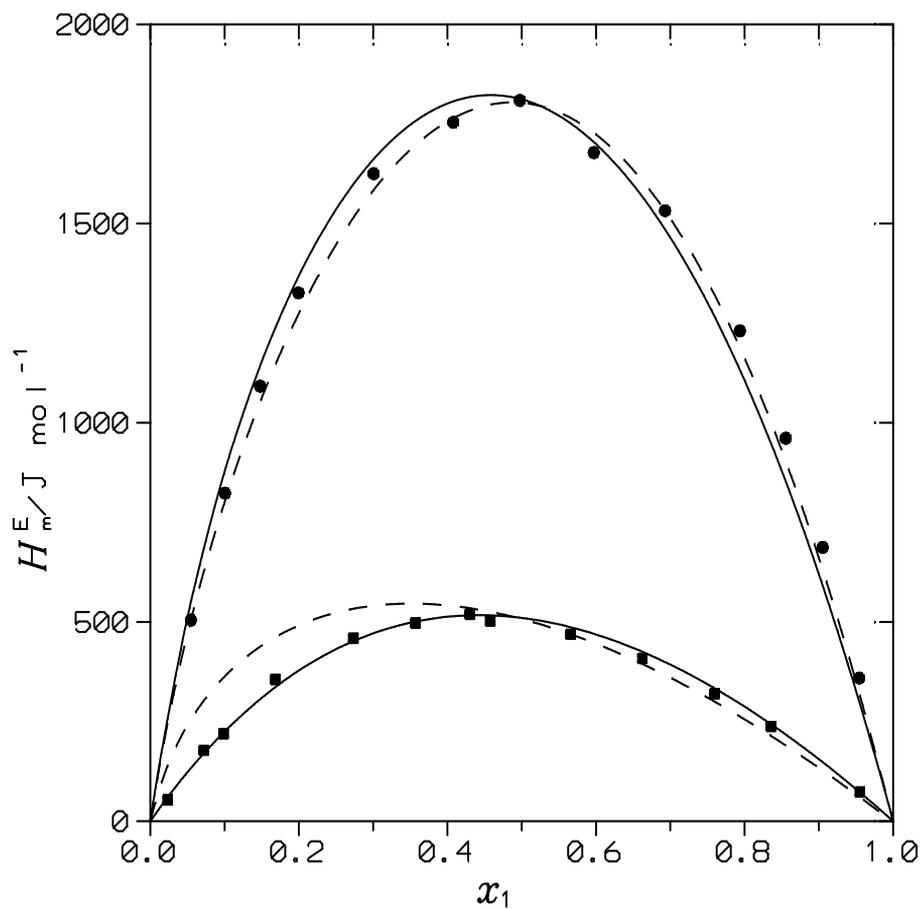

**Figure 1**. $H_m^E$ of benzylamine(1) + hydrocarbon(2) mixtures at temperature $T$ and 0.1 MPa. Points, experimental results: (●), heptane ($T$ = 298.15 K, this work); (■), benzene ($T$ = 303.15 K, [54]). Solid lines, DISQUAC calculations with interaction parameters listed in Table 6. Dashed lines, ERAS results using parameters from Tables 7 and 8.

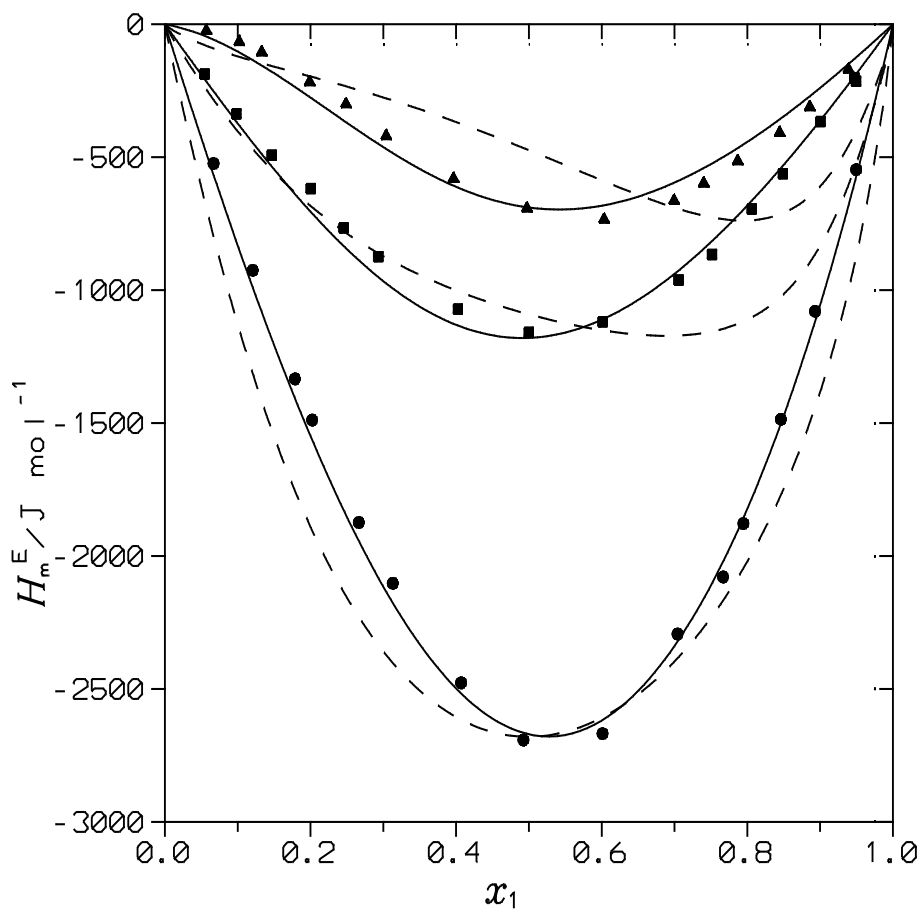

**Figure 2**. $H_m^E$ of 1-alkanol(1) + benzylamine(2) mixtures at 298.15 K and 0.1 MPa. Points, experimental results (this work): (●), methanol; (■), 1-pentanol; (▲), 1-decanol. Solid lines, DISQUAC calculations with interaction parameters listed in Table 6. Dashed lines, ERAS results using parameters from Tables 7 and 8.

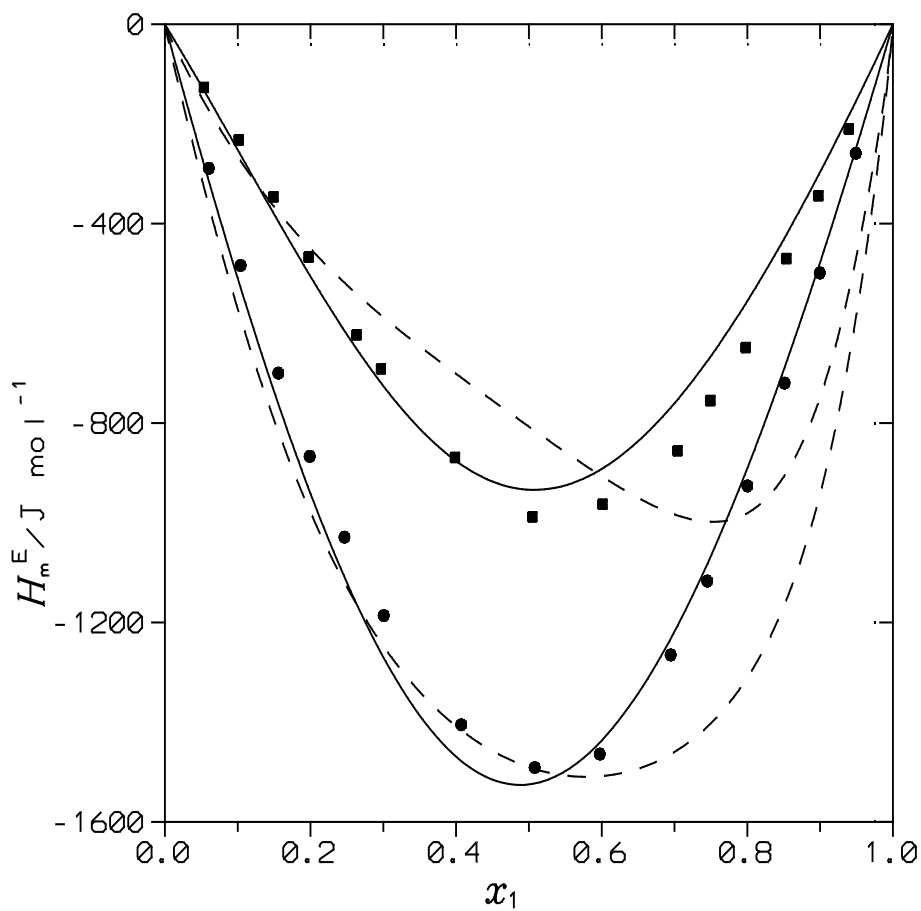

**Figure 3**. $H_m^E$ of 1-alkanol(1) + benzylamine(2) mixtures at 298.15 K and 0.1 MPa. Points, experimental results (this work): (●), 1-propanol; (■), 1-heptanol. Solid lines, DISQUAC calculations with interaction parameters listed in Table 6. Dashed lines, ERAS results using parameters from Tables 7 and 8.

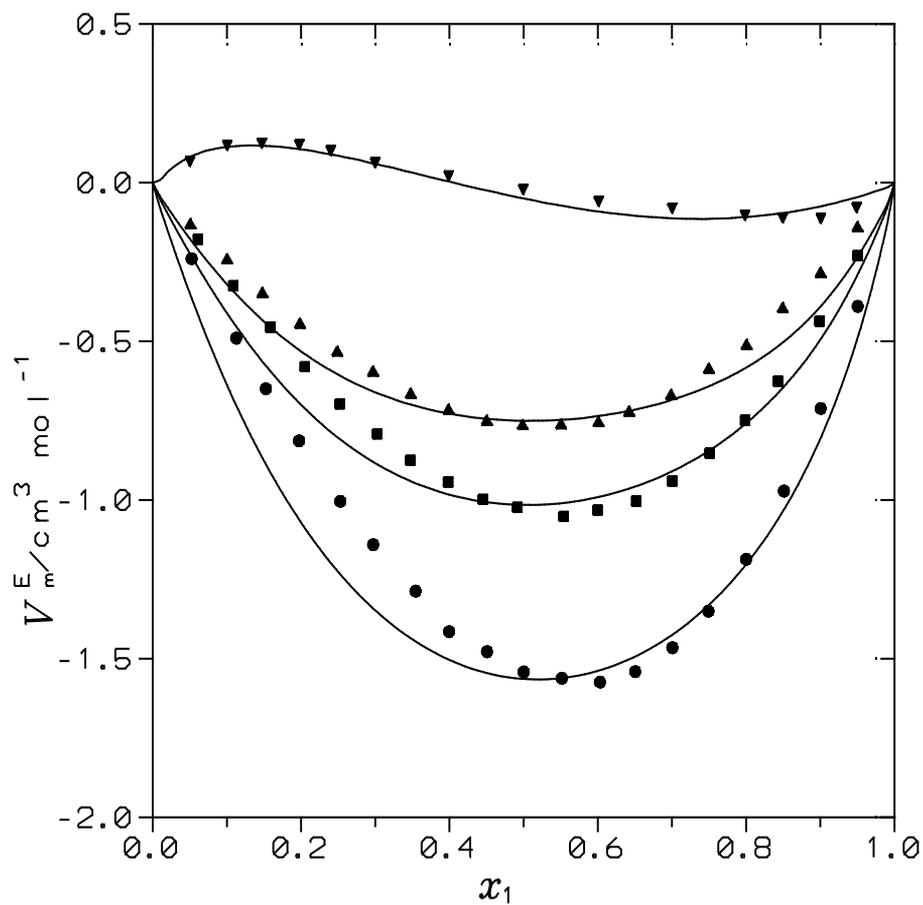

**Figure 4**. $V_m^E$ of mixtures containing benzylamine at 298.15 K and 0.1 MPa. Points experimental results: (▼), benzylamine(1) + heptane(2) (this work); (●), methanol(1) + benzylamine(2); (■), 1-propanol(1) + benzylamine(2); (▲), 1-pentanol(1) + benzylamine(2) [18]. Solid lines, ERAS calculations using parameters listed in Tables 7 and 8.

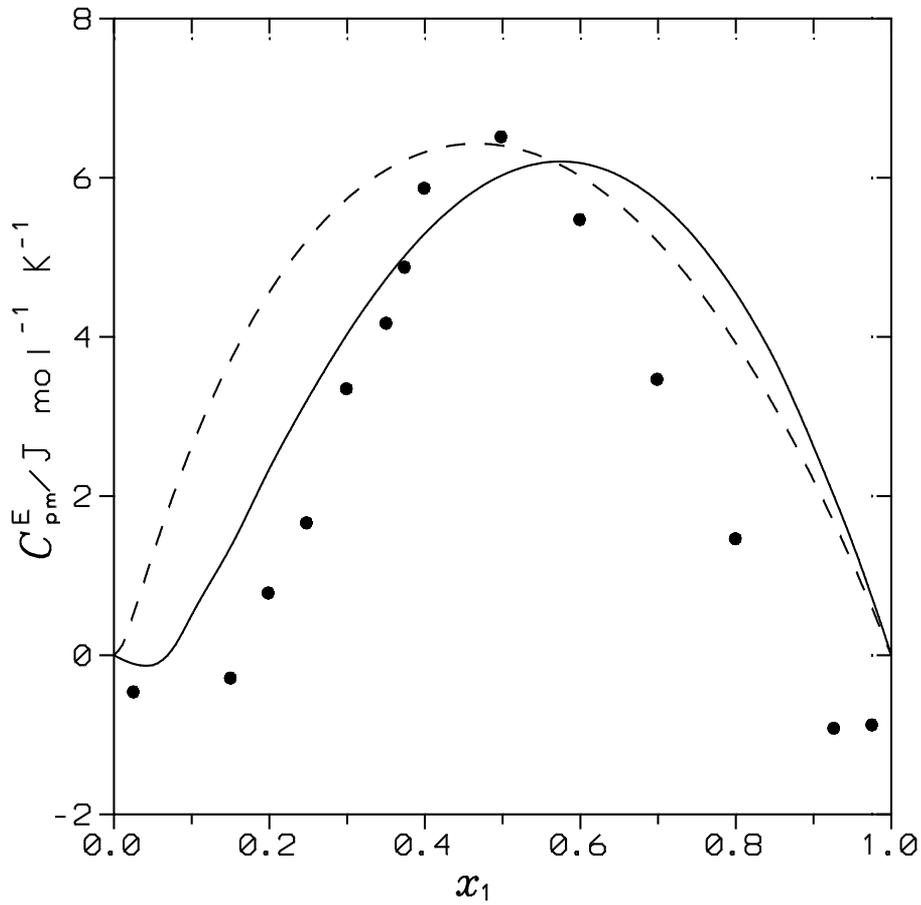

**Figure 5.** $C_{pm}^E$ of the benzylamine(1) + heptane(2) mixture at 293.15 K and and 0.1 MPa. Points, experimental results [19]. Solid line, DISQUAC calculations with interaction parameters listed in Table 6. Dashed line, ERAS results using parameters from Tables 7 and 8 and $\frac{dX_{AB}}{dT} = 0.13$ J cm$^{-3}$ K$^{-1}$.

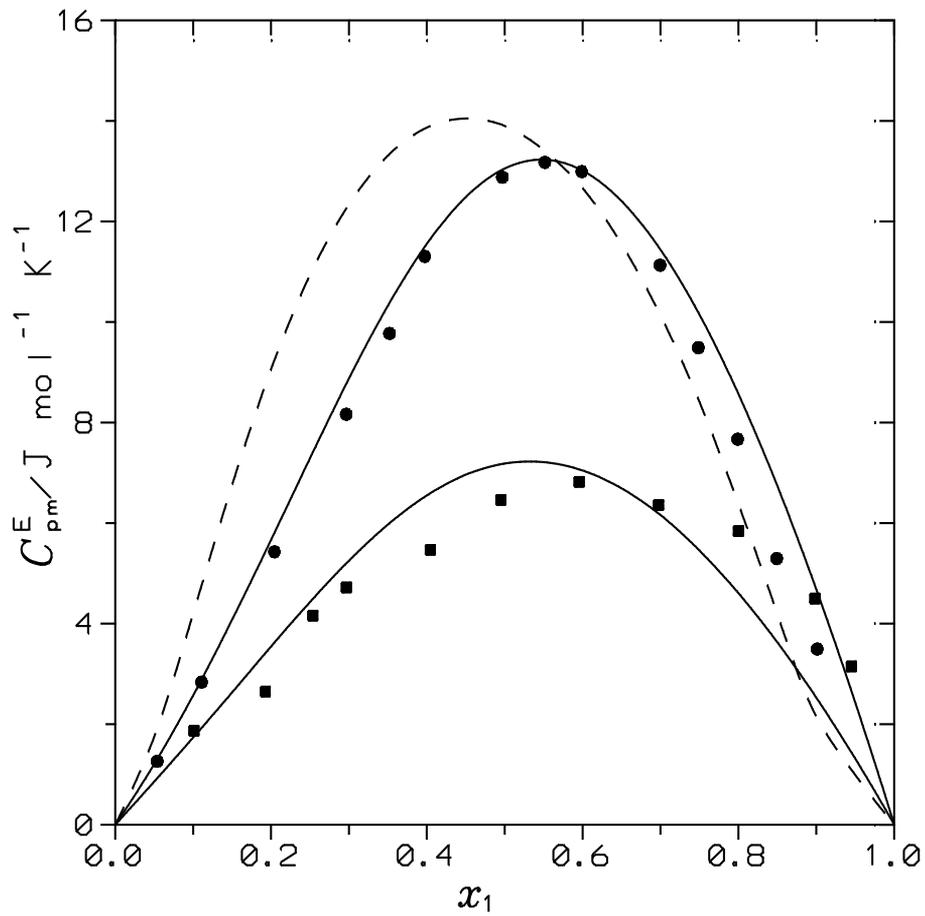

**Figure 6.** $C_{pm}^{E}$ of 1-alkanol(1) + benzylamine(2) mixtures at temperature $T$ and and 0.1 MPa. Points, experimental results [19]: (●), methanol ($T$ = 298.15 K); (■), 1-pentanol ($T$ = 333.15 K). Solid lines, DISQUAC calculations with interaction parameters listed in Table 6. Dashed line, ERAS results for the methanol system using parameters from Tables 7 and 8 and $\frac{dX_{AB}}{dT} = 1$ J cm$^{-3}$ K$^{-1}$.
.

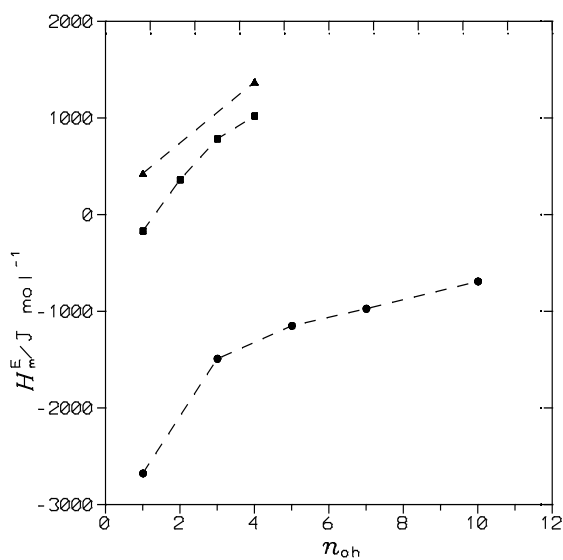 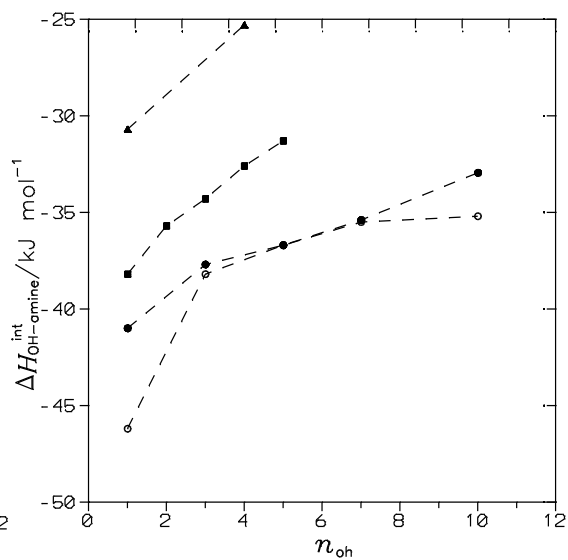

**Figure 7a**                                                **Figure 7b**

**Figure 7**     $H_m^E$ values at equimolar composition, 298.15 K and 0.1 MPa (Figure 7a) and $\Delta H_{OH-NH2}^{int}$ results at 298.15 K and 0.1 MPa (Figure 7b) for 1-alkanol(1) + aromatic amine mixtures(2): (●), benzylamine systems (this work); (■), aniline mixtures [1]; (▲), solutions involving *N*-methylaniline [78]. Open symbols, $\Delta h_{AB}^*$ values from the ERAS model for benzylamine mixtures (this work). Lines, are for the aid of the eye.

SUPPLEMENTARY MATERIAL

# THERMODYNAMICS OF MIXTURES CONTAINING AMINES. XVII. $H_m^E$ and $V_m^E$ MEASUREMENTS FOR BENZYLAMINE + HEPTANE OR + 1-ALKANOL MIXTURES AT 298.15 K. APPLICATION OF THE DISQUAC AND ERAS MODELS

Luis Felipe Sanz, Juan Antonio González,* Fernando Hevia, Isaías. García de la Fuente, and José Carlos Cobos

[a]G.E.T.E.F., Departamento de Física Aplicada, Facultad de Ciencias, Universidad de Valladolid, Paseo de Belén, 7, 47011 Valladolid, Spain.

*corresponding author, e-mail: jagl@termo.uva.es; Fax: +34-983-423136; Tel: +34-983-423757

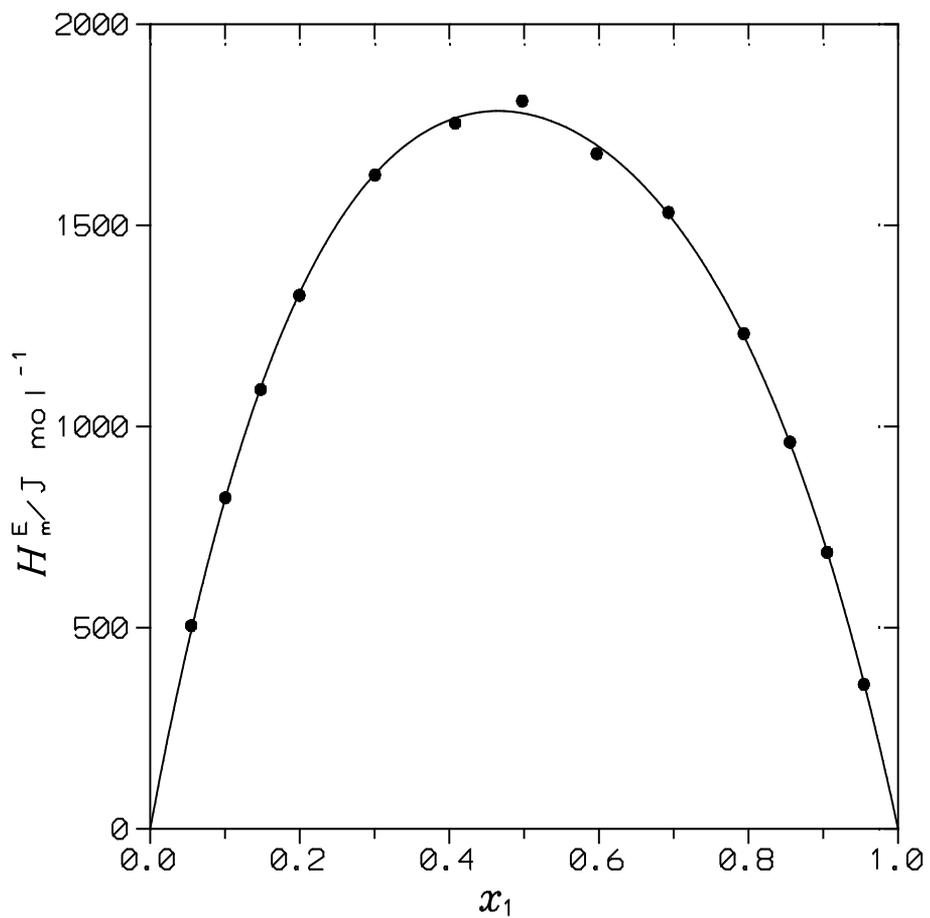

**Figure S1**. $H_m^E$ of the benzylamine(1) + heptane(2) mixture at 298.15 K and 0.1 MPa. Points, experimental values (this work) Solid line, results from equation (1) with parameters listed in Table 4.

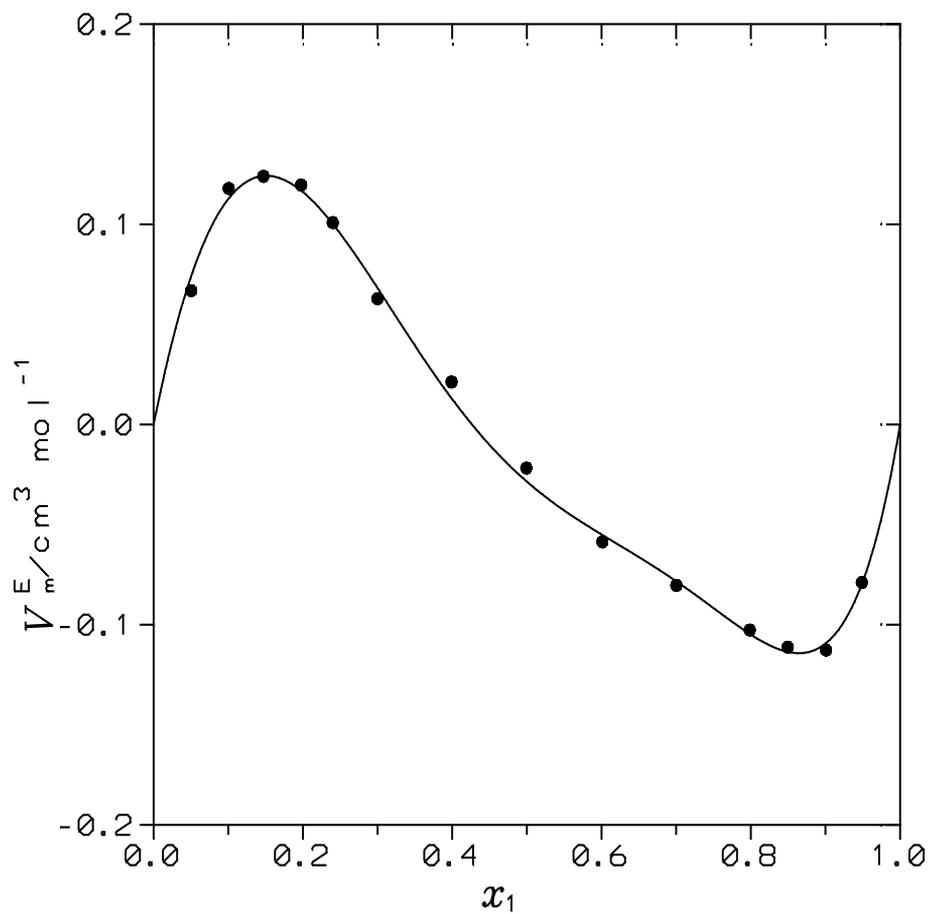

**Figure S2.** $V_m^E$ of the benzylamine(1) + heptane(2) mixture at 298.15 K and 0.1 MPa. Points, experimental values (this work) Solid line, results from equation (1) with parameters listed in Table 4.

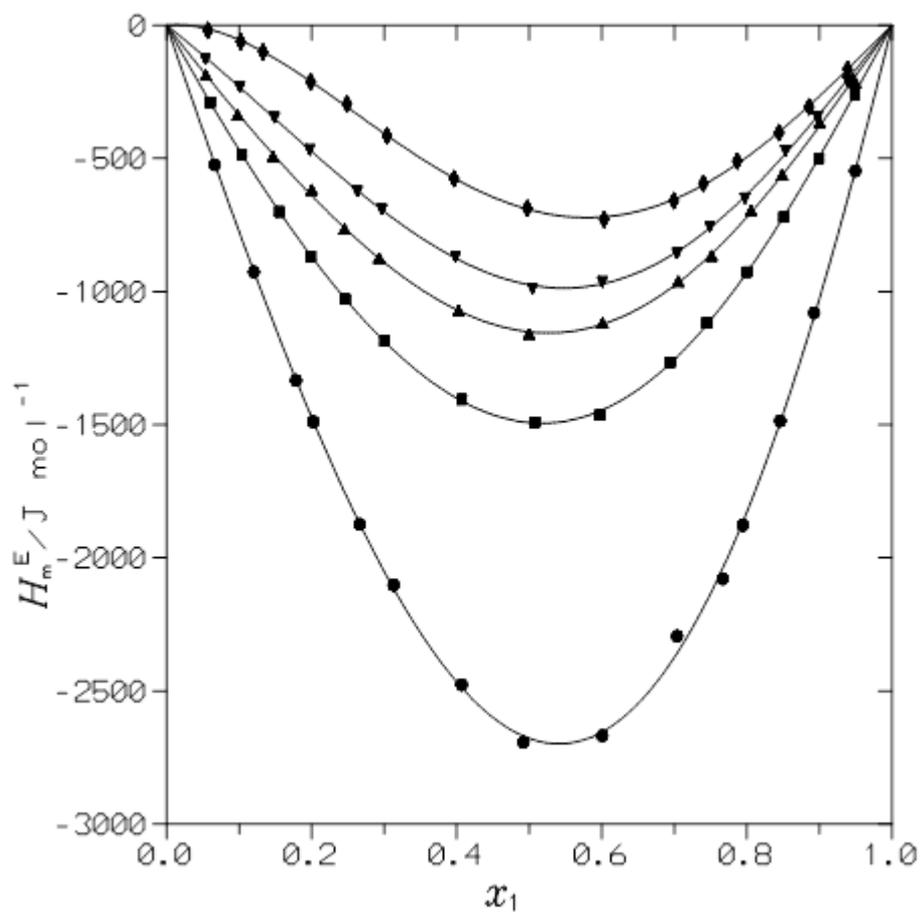

**Figure S3.** $H_m^E$ of 1-alkanol(1) + benzylamine(2) mixtures at temperature 298.15 K and 0.1 MPa. Points, experimental results (this work): (●), methanol; (■), 1-propanol; (▲), 1-pentanol; (▼), 1-heptanol; (◆), 1-decanol. Solid lines, results from equation (1) with parameters listed in Table 5.

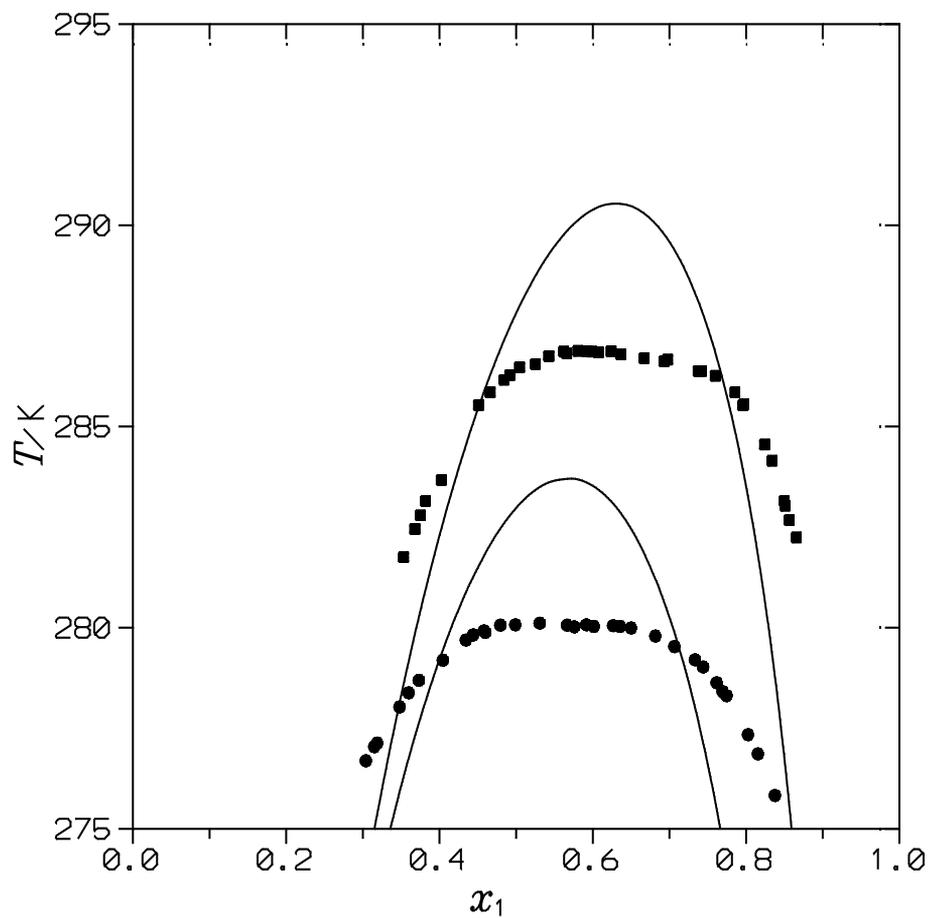

**Figure S4.** LLE of benzylamine(1) + *n*-alkane(2) mixtures. Points, experimental results [55]: (●), dodecane; (■), tetradecane. Solid lines, DISQUAC calculations with interaction parameters from Table 6.

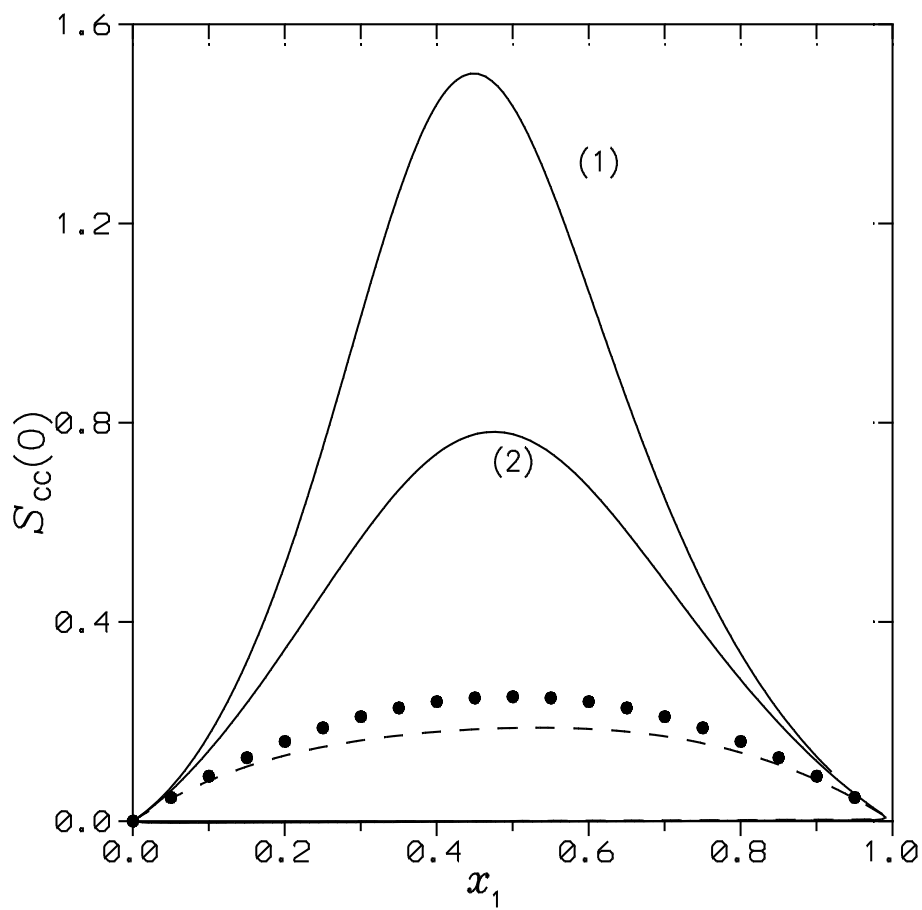

**Figure S5**  DISQUAC calculations on $S_{CC}(0)$ at 298.15 K with interaction parameters from Table 6 for the systems: (1), benzylamine(1) + heptane(2); (2), *N*-methylaniline(1) + heptane(2); dashed line, methanol(1) + benzylamine(2). Points, ideal mixture.